\documentclass{article}

\usepackage{arxiv}
\usepackage{amsmath}
\usepackage[utf8]{inputenc} 
\usepackage[T1]{fontenc}    
\usepackage{hyperref}       
\usepackage{url}            
\usepackage{booktabs}       
\usepackage{amsfonts}       
\usepackage{nicefrac}       
\usepackage{microtype}      
\usepackage{lipsum}
\usepackage{fancyhdr}       
\usepackage{graphicx}       
\graphicspath{{media/}}     

\usepackage{float}
\usepackage{amsmath,amssymb}
\usepackage{graphicx}
\usepackage{booktabs}
\usepackage{setspace}
\usepackage{hyperref}
\usepackage{tfrupee}
\usepackage{threeparttable}
\usepackage{titlesec}
\usepackage{parskip}            

\title{Input--Output Price Parity and Farm Profitability:
A Strategic Perspective for Karnataka
\thanks{\textit{\underline{Citation}}: 
\textbf{Vaishnavi, Lokesha, H., Vedamurthy, K.B., and Patil, M. Input-Output Price Parity and Farm Profitability: A Strategic Perspective for Karnataka. Indian Journal of Economic Development, 21(4): 713--720. DOI: 10.35716/IJED-25143.}}
}

\author{
    Vaishnavi \\
    Assistant Coordinator, MBA (Food Business), Dairy Science College, KVAFSU, Bengaluru, India \\
    \texttt{vaishnavimadari@hotmail.com}
    \And
    Lokesha, H. \\
    Professor, Dept. of Agricultural Economics, University of Agricultural Sciences, Bengaluru, India \\
    \texttt{lokeshakananur@gmail.com}
    \And
    Vedamurthy, K. B. \\
    Associate Professor, Dairy Science College, KVAFSU, Bengaluru, India \\
    \texttt{vedandri@gmail.com}
    \And
    Manojkumar Patil \\
    Research Associate, Indian Institute of Science, Bengaluru, India \\
    \texttt{pmanojkumar@iisc.ac.in}
}

\begin{document}
\maketitle

\begin{abstract}
Agricultural pricing policies are crucial for farm profitability and food security in India. This study analysed how input and output prices significantly influence the profitability of cereals in Karnataka, with the strategic support prices playing a crucial role in maintaining the price parity. The average annual TFP growth was 1.041 \%. Rising input costs, particularly for human labour, led to reduced profitability for Jowar (6.12 \%) and Ragi (4.89 \%). The net effect was adverse for Jowar (-1.50 \%) and Ragi (-0.86 \%) due to rising input costs outpacing output prices. The study recommended increasing the MSP for Jowar (60 \%) and Ragi (46.24 \%) above the existing levels. A strategic price adjusted for changing input costs can stabilise farm incomes and promote sustainable production, enabling efficient pricing policies.
\end{abstract}


\section{Introduction}

Agriculture and allied sectors play a crucial role in the economy of India,
contributing approximately 18 percent to total Gross Value Added (GVA) at
current prices, with an average real GVA growth of 4.3 percent from 2018--19
to 2022--23. However, growth fell to 1.4 percent in 2023--24, primarily due
to extreme El~Niño-induced weather events. Despite the disruptions caused by
the COVID-19 pandemic, the sector demonstrated resilience, recording a 3.7
percent growth in 2020--21. Foodgrain production reached a record 303.30
million tonnes in 2020--21~\cite{des2021karnataka}. However, the agriculture
sector remains highly vulnerable to climatic fluctuations, highlighting the
urgent need for a robust policy framework to ensure price stability and income
security for farmers~\cite{rahman2015asymmetric}. Price factors significantly
influence supply through input and output prices, which directly impact
production costs and profitability~\cite{sadasivam1993price,reddy2009pulses,srivastava2017cost}.

Agricultural pricing policies, particularly the Minimum Support Price (MSP),
play a pivotal role in ensuring economic sustainability for farmers while
maintaining national food security~\cite{chand2003msp}. MSP serves as a key
policy instrument to stabilise farmers' income and influence cropping patterns
in response to market conditions~\cite{sahana2021perception}. However, setting
an effective MSP requires a comprehensive understanding of agricultural cost
structures. Rising input costs, including fertilisers, pesticides, labour, and
machinery, pose significant challenges to farm profitability, making accurate
cost estimation essential for informed policy formulation~\cite{chintapalli2022msp}.

Policy decisions are often influenced by political and economic considerations,
with periodic revisions aimed at enhancing farmers' income and ensuring
equitable price realisation. It is imperative to study the relationships among
factor and product prices and crop profitability, as these are crucial for
determining MSP~\cite{sahana2021perception,parikh2007msp}. Strategic MSP
adjustments can mitigate the impact of inflation and input-output price
volatility, stabilising farm incomes while ensuring sustainable production
levels. This study contributes to the ongoing discourse on MSP effectiveness
by providing empirical evidence on cost structures, market trends, and the
economic implications of pricing interventions.

\section{Methodology}

The present study analysed the impact of input-output prices on the
profitability of cereal crops (Jowar and Ragi) using secondary data for
2019--20 from the Commission for Agricultural Costs and Prices (CACP),
Ministry of Agriculture and Farmers Welfare, Government of India. Jowar and
Ragi were selected as major dryland cereals of Karnataka owing to their
significance in food security, livelihoods, and climate resilience. To assess
growth and development, Total Factor Productivity (TFP) was calculated for
selected crops over 2010--11 to 2021--22.

\subsection{Total Factor Productivity (TFP)}

TFP measures the ratio of aggregate output to aggregate input, commonly
estimated using the growth-accounting method~\cite{coelli2005productivity}.
The T\"ornqvist--Theil index, a widely used approximation of the Divisia
index, was preferred for constructing aggregate output and input indices as it
is exact for a linearly homogeneous translog production function---earning the
`superlative' label from~\cite{diewert1976exact}. In logarithmic form:

\begin{equation}
\ln\!\left(\frac{Y_{jt}}{Y_{j,t-1}}\right)
=\frac{1}{2}\sum_{j}\!\left(S_{jt}+S_{j,t-1}\right)
\ln\!\left(\frac{Y_{jt}}{Y_{j,t-1}}\right)
\tag{Output Index}
\end{equation}

\begin{equation}
\ln\!\left(\frac{X_{it}}{X_{i,t-1}}\right)
=\frac{1}{2}\sum_{i}\!\left(S_{it}+S_{i,t-1}\right)
\ln\!\left(\frac{X_{it}}{X_{i,t-1}}\right)
\tag{Input Index}
\end{equation}

\begin{equation}
\ln\!\left(\frac{\mathrm{TFP}_t}{\mathrm{TFP}_{t-1}}\right)
=\sum_{j}R_j\ln\!\left(\frac{Y_{jt}}{Y_{j,t-1}}\right)
-\sum_{i}S_i\ln\!\left(\frac{X_{it}}{X_{i,t-1}}\right)
\tag{TFP Index}
\end{equation}

\noindent where $Y_{jt}$ is output $j$ at time $t$; $X_{it}$ is input $i$ at
time $t$; $R_j$ is the elasticity of output with respect to $j$; $S_{it}$ and
$S_{jt}$ are the shares of input $i$ in total cost and output $j$ in total
revenue, respectively.

A chain index was preferred over a fixed-base index~\cite{coelli2005productivity,kannan2017tfp}:
\begin{equation}
I(0,t)=I(0,1)\times I(1,2)\times I(2,3)\times\cdots\times I(t-1,t)
\end{equation}

\subsection{Translog Cost Function}

For minimisation of total cost $C$, subject to a production function, the
minimum cost function is $C^{*}=f(Q,p_1,p_2,\dots,p_n)$, where $Q$ is total
output and $p_i$ is the price of the $i^{\text{th}}$ input. Its logarithmic
Taylor series expansion is~\cite{kumar2010factor,srivastava2017cost}:

\begin{equation}
\ln C(w,y)=\alpha_0+\sum_{i}\alpha_i\ln w_i+\alpha_y\ln y
+\tfrac{1}{2}\sum_{i}\sum_{j}\alpha_{ij}\ln w_i\ln w_j
+\tfrac{1}{2}\alpha_{yy}(\ln y)^2+\sum_{i}\alpha_{iy}\ln w_i\ln y
\end{equation}

Using Shephard's lemma, the derived cost-share equations are:
\begin{equation}
S_i=\alpha_i+\alpha_{iy}\ln y+\sum_{j}\alpha_{ij}\ln w_j,
\qquad S_i=\frac{w_i x_i}{C}
\end{equation}

Symmetry ($\alpha_{ij}=\alpha_{ji}$) and homogeneity of degree one in prices
impose:
\begin{equation}
\sum_{i}\alpha_i=1,\quad\sum_{j}\alpha_{ij}=0,\quad\sum_{i}\alpha_{iy}=0
\end{equation}

The Allen partial elasticities of substitution and own/cross-price elasticities
are:
\begin{align}
\sigma_{ii}&=\frac{\alpha_{ii}+S_i^2-S_i}{S_i^2}, &
\sigma_{ij}&=\frac{\alpha_{ij}+S_iS_j}{S_iS_j}\;(i\neq j), &
\eta_{ij}&=\sigma_{ij}S_j
\end{align}

\subsection{Strategic Support Price}

The Strategic Support Price (SSP) accounts for both input and output price
dynamics and is given by:
\begin{equation}
\text{SSP}=\text{MSP}\times(1-\text{Net Effect})
\end{equation}

The growth in unit price of factor or product $i$ between 2015--16 and
2019--20 is $\Delta P^i=(P^i_{2019\text{-}20}-P^i_{2015\text{-}16})/P^i_{2015\text{-}16}$.
The contribution to supply is $C^i=\Delta P^i\times E^i$ (where $E^i$ is the
supply elasticity), and the net effect is:
\begin{equation}
\text{Net Effect}=\sum_{i}C^i
\end{equation}

To ensure policy realism, the net effect is bounded within $[-2.0,\,0.8]$,
which restricts the SSP to between 20\% and 300\% of the CACP-recommended MSP.

\section{Results and Discussion}

\subsection{Total Factor Productivity for Selected Crops in Karnataka}

Agricultural productivity growth is crucial for both the sector and the
economy. TFP growth in agriculture is driven by improved inputs, technology,
and better farm management. Annual growth trends for selected crops in
Karnataka from 2010--11 to 2021--22 are presented in Table~\ref{tab:tfp_growth}.

\begin{table}[htbp]
\centering\small
\caption{Annual growth in input, output and TFP index of selected crops in
         Karnataka, 2010--11 to 2021--22}
\label{tab:tfp_growth}
\begin{tabular}{lccc}
\toprule
\textbf{Year} & \textbf{Total Input Index} & \textbf{Total Output Index} & \textbf{TFP} \\
\midrule
2010 & 1.00 & 1.00 & 1.00 \\
2011 & 0.96 & 1.06 & 1.10 \\
2012 & 0.99 & 1.10 & 1.11 \\
2013 & 0.89 & 0.79 & 0.89 \\
2014 & 0.97 & 1.27 & 1.30 \\
2015 & 1.05 & 0.80 & 0.77 \\
2016 & 0.82 & 0.84 & 1.02 \\
2017 & 0.81 & 1.03 & 1.28 \\
2018 & 1.41 & 1.21 & 0.86 \\
2019 & 0.84 & 0.89 & 1.05 \\
2020 & 1.07 & 1.04 & 0.97 \\
2021 & 0.98 & 1.13 & 1.15 \\
\midrule
\textbf{Average} & \textbf{0.98} & \textbf{1.01} & \textbf{1.04} \\
\bottomrule
\end{tabular}
\end{table}

The average annual TFP growth was 1.04 percent, reflecting improved input
efficiency driven by technology and better management. Total input growth
stood at 0.98 percent, indicating reduced input use due to limited credit,
adverse weather and other challenges, while total output grew at 1.01 percent
over the period. \cite{goldar2023tfp} emphasised the role of capital
deepening, labour skills, ICT, R\&D, and trade in boosting TFP growth in
India; \cite{kannan2017tfp} linked crop output growth to TFP and
irrigation-driven area expansion, aligning with \cite{kale2021tur}.

The input cost distribution for selected crops in Karnataka (2019--20) showed
that human labour accounts for the largest share (48 percent), followed by
machine labour (16 percent), animal labour (12 percent), fertilisers
(12 percent), seeds (8 percent), plant protection chemicals (3 percent), and
manure (2 percent). Rising input costs are primarily driven by price inflation
rather than increased usage~\cite{choudhary2022adoption} (Table~\ref{tab:input_share}).

\begin{table}[htbp]
\centering\small
\caption{Share of inputs in the total variable costs of selected crops in Karnataka (2019--20)}
\label{tab:input_share}
\begin{tabular}{lc}
\toprule
\textbf{Component} & \textbf{Share (\%)} \\
\midrule
Human Labour   & 48 \\
Machine Labour & 16 \\
Animal Labour  & 12 \\
Fertiliser     & 12 \\
Seed           & 8  \\
PPC            & 3  \\
Manure         & 2  \\
\midrule
\textbf{Total} & \textbf{100} \\
\bottomrule
\end{tabular}
\end{table}

\subsection{Supply Response}

The translog cost function was employed to analyse crop supply responses,
focusing on how input-output price parity affected profitability. Demand
equations for human labour, fertiliser, and machine labour were jointly
estimated with the cost function. Input demand elasticity estimates are
presented in Tables~\ref{tab:jowar_elasticity} and~\ref{tab:ragi_elasticity}.

\begin{table}[htbp]
\centering\small
\caption{Derived estimates of own and cross-price elasticities of input demand for Jowar}
\label{tab:jowar_elasticity}
\begin{tabular}{lcccc}
\toprule
\textbf{Factor} & \textbf{Human Labour} & \textbf{Fertiliser} & \textbf{Machine Labour} & \textbf{Output} \\
\midrule
Human Labour   & $-6.122$ & $3.452$  & $2.670$  & $0.266$ \\
Fertiliser     & $0.660$  & $-0.643$ & $-0.039$ & $0.262$ \\
Machine Labour & $9.685$  & $-0.795$ & $-4.891$ & $0.646$ \\
\bottomrule
\end{tabular}
\end{table}

The negative own-price elasticity of human labour ($-6.122$) indicates
that a 1 percent rise in labour costs leads to a significant decline in
demand for jowar production. Fertiliser and machine labour elasticities
($-0.643$ and $-4.891$, respectively) also reflect inelastic demand.
Positive output price elasticities confirm that higher jowar prices encourage
greater input use, consistent with \cite{srivastava2017cost} and
\cite{ashrit2021input}.

\begin{table}[htbp]
\centering\small
\caption{Derived estimates of own and cross-price elasticities of input demand for Ragi}
\label{tab:ragi_elasticity}
\begin{tabular}{lcccc}
\toprule
\textbf{Factor} & \textbf{Human Labour} & \textbf{Fertiliser} & \textbf{Machine Labour} & \textbf{Output} \\
\midrule
Human Labour   & $-2.279$ & $1.261$  & $1.018$  & $-1.224$ \\
Fertiliser     & $1.189$  & $-0.549$ & $-0.640$ & $0.649$  \\
Machine Labour & $2.022$  & $-1.349$ & $-0.674$ & $0.036$  \\
\bottomrule
\end{tabular}
\end{table}

The negative own-price elasticities for human labour ($-2.279$), fertiliser
($-0.549$), and machine labour ($-0.674$) in ragi production indicate that
price increases significantly reduced input use. The cross-price elasticity
for human labour with output ($-1.224$) underscores ragi's labour-intensive
nature, consistent with findings of \cite{srivastava2017cost}.

\begin{table}[htbp]
\centering\small
\caption{Impact of factor and product prices on the profit of Jowar}
\label{tab:jowar_profit}
\begin{tabular}{lccccc}
\toprule
\textbf{Factor/Product} &
\multicolumn{3}{c}{\textbf{Unit Price}} &
\textbf{Elasticity} &
\textbf{Contribution (\%)} \\
\cmidrule(lr){2-4}
 & \textbf{2015--16} & \textbf{2019--20} & \textbf{Growth} & & \\
\midrule
Output         & 5.10   & 11.28  & 1.21 & $0.65$  & $0.78$  \\
Labour         & 28.32  & 31.22  & 0.10 & $-6.12$ & $-0.63$ \\
Fertiliser     & 37.03  & 38.35  & 0.04 & $-0.64$ & $-0.02$ \\
Machine Labour & 265.39 & 354.16 & 0.33 & $-4.89$ & $-1.64$ \\
\midrule
\textbf{Net Effect} & \multicolumn{5}{c}{$\mathbf{-1.50}$} \\
\bottomrule
\end{tabular}
\end{table}

Between 2015--16 and 2019--20, the unit price of jowar rose from 5.10 to
11.28 (positive elasticity 0.65), yet a 10 percent rise in labour costs
($\eta=-6.12$) and a 33 percent rise in machine labour costs eroded
profitability. The overall net effect of $-1.50$ indicates that combined
input cost pressures outweighed output price gains (Table~\ref{tab:jowar_profit}).

\begin{table}[htbp]
\centering\small
\caption{Impact of factor and product prices on the profit of Ragi}
\label{tab:ragi_profit}
\begin{tabular}{lccccc}
\toprule
\textbf{Factor/Product} &
\multicolumn{3}{c}{\textbf{Unit Price}} &
\textbf{Elasticity} &
\textbf{Contribution (\%)} \\
\cmidrule(lr){2-4}
 & \textbf{2015--16} & \textbf{2019--20} & \textbf{Growth} & & \\
\midrule
Output         & 19.10  & 18.11  & $-0.05$ & $0.04$  & $0.00$  \\
Labour         & 41.45  & 39.87  & $-0.04$ & $-2.28$ & $0.09$  \\
Fertiliser     & 33.02  & 31.81  & $-0.04$ & $-0.55$ & $0.02$  \\
Machine Labour & 245.28 & 595.09 & $1.43$  & $-0.67$ & $-0.96$ \\
\midrule
\textbf{Net Effect} & \multicolumn{5}{c}{$\mathbf{-0.86}$} \\
\bottomrule
\end{tabular}
\end{table}

For ragi, a 5 percent fall in output prices and a 143 percent surge in
machine labour costs drove a net effect of $-0.86$ percent, indicating
that rising production costs outpaced output prices (Table~\ref{tab:ragi_profit}).

\subsection{Determination of Strategic Support Price}

The MSP recommended by the CACP comprises only the $A2+FL$ cost. The
Strategic Support Price (SSP) additionally accounts for total cost,
market prices, and demand-supply conditions, offering farmers a more
accurate estimate of production costs (Table~\ref{tab:msp_methods}).

\begin{table}[H]
\centering\small
\begin{threeparttable}
\caption{Determination of MSP through different pricing methods for selected crops (2019--20)}
\label{tab:msp_methods}
\begin{tabular}{lcc}
\toprule
\textbf{Particulars} & \textbf{Jowar} & \textbf{Ragi} \\
\midrule
Cost A2+FL (\rupee/qtl)                                    & 3{,}043 & 3{,}557 \\
Cost C2 (\rupee/qtl)                                       & 3{,}143 & 3{,}646 \\
MSP recommended by CACP (\rupee/qtl)                       & 2{,}550 & 3{,}150 \\
Swaminathan MSP (C2+50\%) (\rupee/qtl)                     & 4{,}715 & 5{,}469 \\
Gap: CACP vs.\ Swaminathan (\%)                            & 45.92  & 42.40  \\
Strategic Support Price (\rupee/qtl)                        & 6{,}375 & 5{,}859 \\
Gap: CACP MSP vs.\ SSP (\%)                                & 60.00  & 46.24  \\
\bottomrule
\end{tabular}
\begin{tablenotes}
\footnotesize
\item Percentages use the target price as base to show the policy gap.
\end{tablenotes}
\end{threeparttable}
\end{table}

During 2019--20, the gap between the CACP-recommended MSP and the Swaminathan
Commission's C2+50\% price was 45.92 percent for jowar and 42.40 percent for
ragi. The SSP framework further widens this gap to 60 percent and 46.24
percent, respectively, reflecting the full burden of input cost inflation
that current MSP policy does not address.

\section{Conclusion}

This study provided empirical evidence on factors influencing crop prices
in Karnataka, using the translog cost model to estimate input demand
elasticities and inform MSP formulation. Stabilising key input costs---
fertilisers, labour, and machinery---is crucial for managing overall
production expenses. Accurate cost estimation enables targeted subsidies,
supports output-based incentives, and promotes agricultural growth. The
Strategic Support Price framework, by adjusting the MSP to reflect actual
input cost dynamics, can serve as an income stabiliser for farmers and
ensure sustainable agricultural growth. Future research could analyse
state-wise variations using farm-level data to capture technology adoption,
cost dynamics, and productivity trends, offering practical insights to
refine and regionalise MSP policies.

\bibliographystyle{unsrt}  
\bibliography{references}

\end{document}